\documentclass{ecai} 

\usepackage{latexsym}
\usepackage{amssymb}
\usepackage{amsmath}
\usepackage{amsthm}
\usepackage{booktabs}
\usepackage{enumitem}
\usepackage{graphicx}
\usepackage{color}
\usepackage{float}

\newcommand{\BibTeX}{B\kern-.05em{\sc i\kern-.025em b}\kern-.08em\TeX}

\begin{document}


\begin{frontmatter}


\paperid{123} 


\title{Influence of prior and task generated emotions on XAI explanation retention and understanding}


\author[A,B]{\fnms{Birte}~\snm{Richter}\orcid{0000-0002-0957-2406}\thanks{Corresponding Author. Email: birte.richter@uni-bielefeld.de}}
\author[A,B]{\fnms{Christian}~\snm{Schütze}\orcid{0000-0002-8860-0478}}
\author[A]{\fnms{Anna}~\snm{Aksonova}\orcid{0009-0002-3740-8555
}} 
\author[A,B]{\fnms{Britta}~\snm{Wrede}\orcid{0000-0003-1424-472X}} 
\address[A]{Medical Assistance Systems, Medical School OWL}
\address[B]{Center for Cognitive Interaction Technology (CITEC), Bielefeld University}


\begin{abstract}
The explanation of AI results and how they are received by users is an increasingly active research field. However, there is a surprising lack of knowledge about how social factors such as emotions affect the process of explanation by a decision support system (DSS). 
While previous research has shown effects of emotions on DSS supported decision-making, it remains unknown in how far emotions affect cognitive processing during an explanation.
In this study, we, therefore, investigated the influence of prior emotions and task-related arousal on the retention and understanding of explained feature relevance. 
To investigate the influence of prior emotions, we induced happiness and fear prior to the decision support interaction. Before emotion induction, user characteristics to assess their risk type were collected via a questionnaire.
To identify emotional reactions to the explanations of the relevance of different features, we observed heart rate variability (HRV), facial expressions, and self-reported emotions of the explainee while they were observing and listening to the explanation and assessed their retention of the features as well as their understanding of the influence of each feature on the outcome of the decision task. 
Results indicate that (1) task-unrelated prior emotions do not affected the retention  but may affect the understanding of the relevance of certain features in the sense of an emotion-induced confirmation bias, (2) certain features related to personal attitudes yielded arousal in individual participants, (3) this arousal affected the understanding of these variables.
\end{abstract}

\end{frontmatter}


\section{Introduction and Related Work}

As artificial intelligence (AI) systems increasingly support human decision-making across diverse domains—from healthcare to finance and beyond—there is a growing demand for these systems to be not only accurate, but also explainable. 
Explainable AI (XAI) aims to make machine-generated decisions transparent and interpretable, allowing users to understand and potentially trust the reasoning behind automated recommendations. A key goal of XAI research is to ensure that users can recall and understand the explanations provided by decision support systems (DSSs), particularly when those explanations concern the relevance of specific input features.

While early work has focused on providing mathematical explanations for AI researchers themselves, lay users as well as domain experts (i.e., medical experts) have been identified as an important target group. This shift in focus has lead to a re-evaluation of existing research, identifying the need for more interactive approaches \cite{rohlfing2020explanation, schmid2022ki}.
However, the underlying assumption in this research has mostly been that interaction takes place with a rational decision maker who follows purely logical considerations. Thus, support has been intended to (1) provide the human decision maker with relevance information about certain features, and (2) to avoid cognitive biases such as confirmation bias \cite{wang2019designing, battefeld2024revealing}.
Yet, it is well known that human decision-making is heavily influenced by emotions \cite{george2016affect}. 
\begin{figure}[t!]
    \centering
    \includegraphics[width=0.5\linewidth]{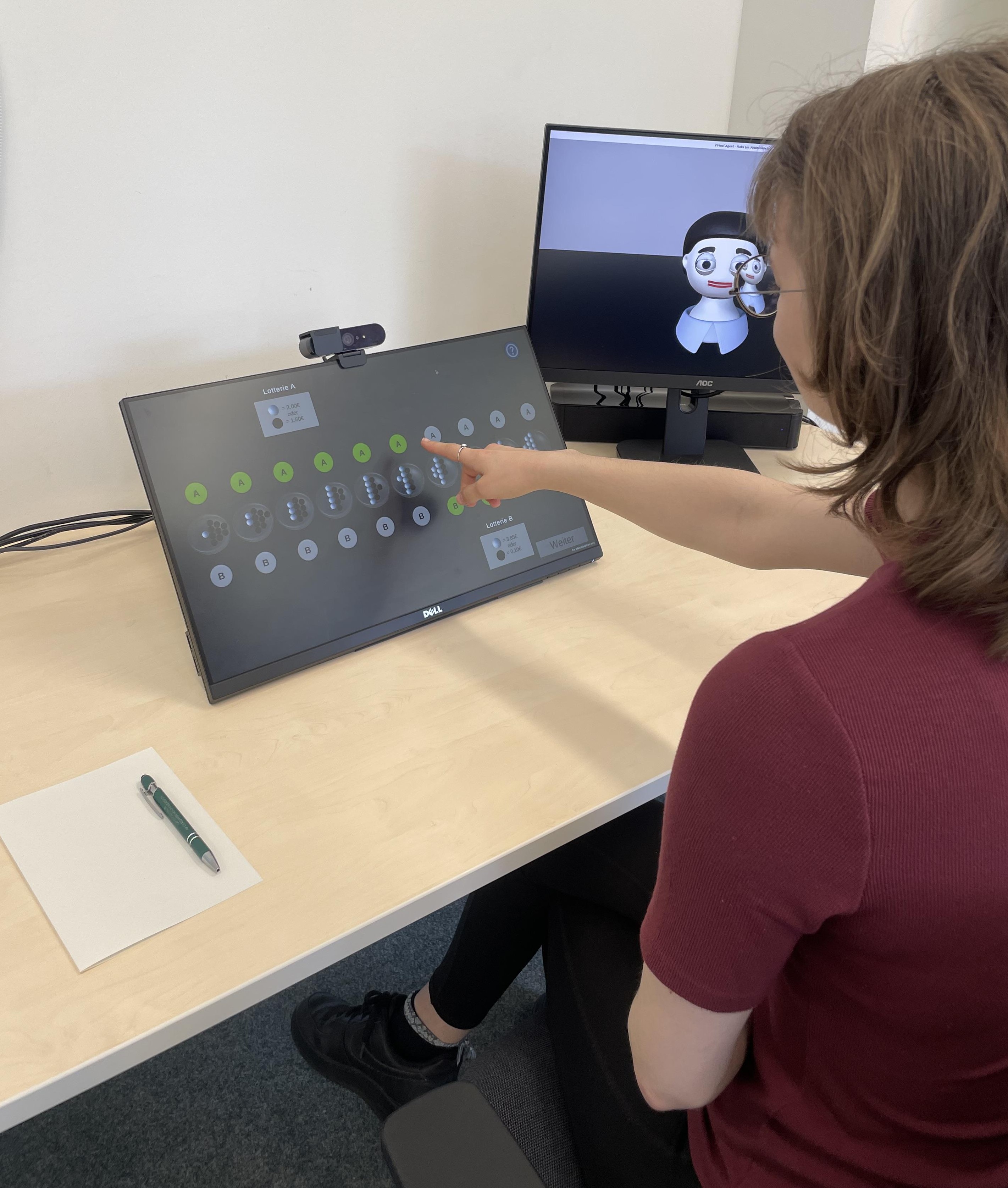}
    \caption{Study Setup during the first Hold and Laury decission (staged)}
    \label{fig:setup}
\end{figure}

More recently, the influence of emotions on the outcome of AI explanations has come into the focus of research. For example, it was shown that task-unrelated prior emotions affected advice taking behavior of participants given different explanation strategies \cite{thommes2024xai}. More specifically, participants with high arousal were more likely to follow AI advice with a (guided) explanation than participants with low arousal. The latter preferred AI advice without any explanation.

On the other hand, explanations can induce affective reactions. Explanations given by an AI for an easy task were shown to yield negative effect in explainees whereas explanations in a difficult task caused positive affect. But also the quality of advice can affect the emotional state of an explainee. In a vignette study it was found that wrong advice would lead to negative feelings whereas correct advice would lead to positive feelings \cite{bernardo2022iscid}. 

Interestingly, an investigation on the effectiveness of an XAI intervention showed that an explicit nudging strategy was successful in debiasing emotions \cite{lammert2025}. However, it was not strong enough to debias decision making.

However, clear guidelines regarding how emotions affect the understanding in a decision-making situation and how an explaining system should take the explainee's emotional state into account are still missing. This work contributes to the field by addressing this question.




\section{Research Questions}

It is important to note that emotions can arise as contextual factors, i.e., from unrelated tasks, or from the interaction with the system concerning the task at hand. While both emotions are likely to have influence on the human's processing, it is important to separate the effects of prior task-unrelated emotions from those that are closely related with the task at hand.

In this research, we therefore investigate three research questions:
\begin{itemize}
    \item \textbf{RQ1a:} Do task-unrelated emotions influence the retention of explained feature relevance?
    \item \textbf{RQ1b:} Do task-unrelated emotions influence the understanding of explained feature relevance?
    \item \textbf{RQ2:} Which features trigger emotional reactions during explanation? (task-related emotions)
    \item \textbf{RQ3a:} Do task-related emotions influence retention of the explanation features?
    \item \textbf{RQ3b:} Do task-related emotions influence understanding of the explanation features?
\end{itemize}

\section{Method}
To investigate the influence of emotions on retention and understanding of explanations, we conducted an interaction study with a decision support system.


\subsection{Measurements}

\subsubsection{Independent Variables}
Two types of emotional influences were considered: (1) \textit{prior task-unrelated emotions}and (2) \textit{task-related emotions}, defined as emotional responses elicited directly by the explanation itself.

\textbf{Prior task-unrelated emotions.}
The prior task-unrelated emotions are induced before the explanation and unrelated to its content.
We chose a between-subjects design, with participants assigned to either a \textit{fear} or a \textit{happiness} condition, as an induced emotion. The emotion induction itself is described in section \ref{sec:ei}.

\textbf{Task-related emotions.}
We measure the emotional reactions by using the EmoNet \cite{gerczuk2021tac} arousal data during the feature presentation and explanation.
Based on the arousal value $a$, an anomaly (emotional reaction) in the arousal state is detected by 
\begin{equation}
    z_{score} = \frac{a - roll_{mean}}{roll_{sd}}
\end{equation}
in combination with the rolling z‑score with k = 2.5 and a window of 500ms. 
\begin{equation}
emotional_{reaction} =
\begin{cases}
1, & \text{if } z > 2.5 \\
0, & \text{else}
\end{cases}
\end{equation}

\subsubsection{Dependent Variables}
\label{sec:DepVar}

\textbf{Retention.}
Retention was measured based on the participant's verbal recall of the features they remembered as relevant to the AI's decision, as conveyed through the explanation. Participants were explicitly asked to provide verbal input, allowing them to articulate their reasoning processes. The spoken responses were automatically transcribed using Whisper, a deep neural network-based automatic speech recognition system. The recognized words were then manually mapped to the ten predefined variable names, accounting for minor inaccuracies in naming.

\textbf{Understanding.}
To assess the extent to which participants understood the meaning of the variables, they were asked—via a graphical user interface (see Fig. \ref{fig:GUIRecallQuestion})—to indicate the contribution of each variable to the overall decision of the decision support system (DSS). Specifically, participants were instructed to indicate whether a given variable contributed to a higher or lower risk estimate. This was implemented by allowing users to move each variable to the left (indicating lower risk) or to the right (indicating higher risk).
We interpreted this recalled information as a cue for (retained) understanding of the explanation.

\subsubsection{Control Variables and Additional Measurements}
In addition to the primary measures, several supplementary variables were recorded for exploratory and control purposes:

\begin{itemize}
    \item Gender: Participant gender (male/female/divers?) was recorded.
    \item STAI:  State-Trait Anxiety Inventory (STAI), providing a measure of participants' baseline anxiety disposition.
    \item mDES: \cite{galanakis2016reliability} modified Differential Emotions Scale, 
    \item SAM \cite{bradley1994measuring}: Self-assessment manekin, providing a measure of participants' self rated arousal and valence
    \item Hear Rate Variability (HRV)
    \item EmoNet \cite{gerczuk2021tac} Results: In addition to arousal scores, full output vectors from the EmoNet model were stored for each facial frame, enabling detailed emotional state tracking over time.

    \item System Events: System-level events (e.g., start of an explanation) were logged for quality control and alignment of multimodal data streams.

    \item Videos: All participant sessions were video recorded for potential qualitative analysis and cross-validation of facial expression data.
\end{itemize}

\subsection{Decision Suport System}
In this study the same decision task and advice scheme was utilized as in a prior study \cite{schuetze2023ichci}.
The system centers around an embodied conversational agent named Flobi, who provides a personalized assessment of an individual's risk profile based on a set of predefined input features. This risk assessment is subsequently used in the context of a Holt and Laury lottery task, where participants make incentivized decisions under risk. The agent's evaluation serves as a form of decision support, offering guidance while allowing participants to ultimately make their own choices.

\subsection{Participants}
A total of 24 participants took part in the study. Of these, 15 were assigned to the fear condition and 9 to the happy condition. The sample included 8 male and 16 female participants.
All participants were students at Bielefeld University.

\subsection{Experimental Setting — Procedure}

\begin{figure}[H]
    \centering
    \includegraphics[width=1\linewidth]{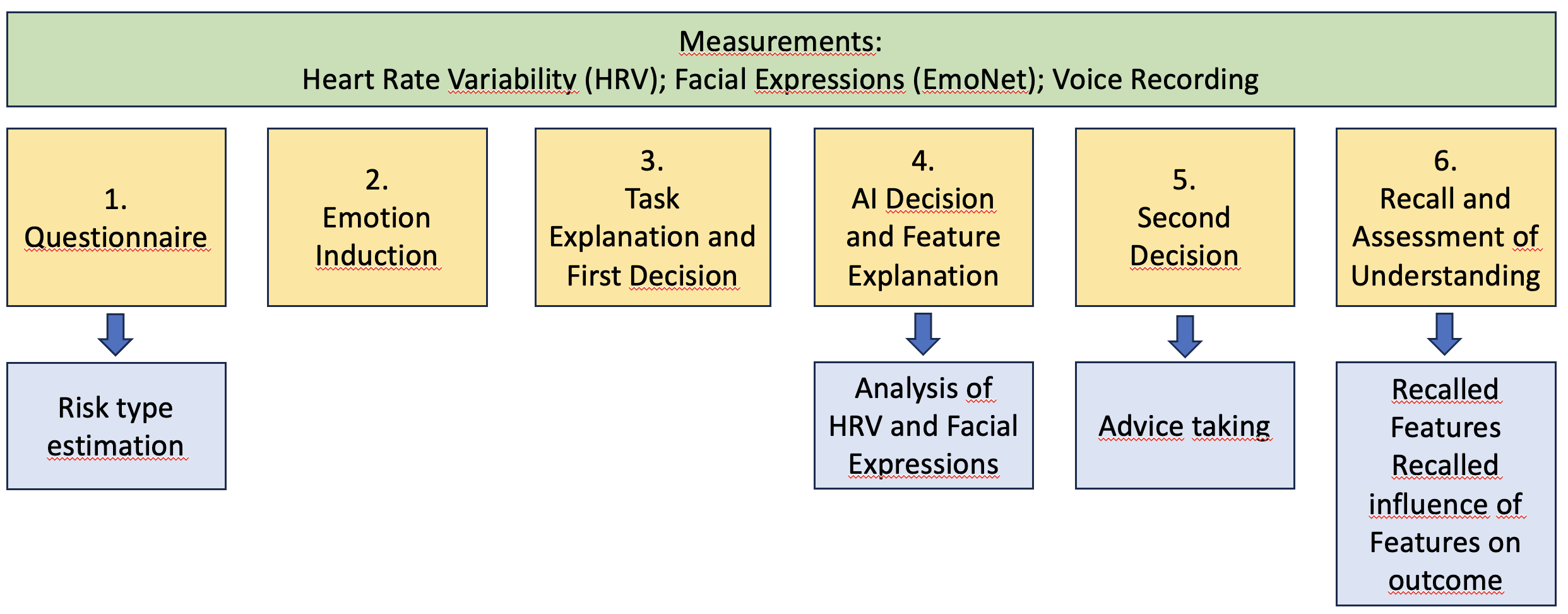}
    \label{fig:procedure}
    \caption{Experimental procedure with six phases. }
\end{figure}

The experiment consisted of six phases (cf. Fig. \ref{fig:procedure}. We recorded the users' heart rate variability, video and audio, and their facial expressions as computed by EmoNet.
In the first phase, a questionnaire was administered that requested data from the user to assess the user's risk type. The risk type classification was achieved by a linear scoring scheme integrating the user's numerical answers, yielding a risk type value between 0 and 9. After an emotion induction sequence, the actual risk task was explained to the user and the user could provide his/her risk selection, i.e., selecting a high or low risk on a scale from 0 to 9.
After the user's first risk decision, the system would present its risk suggestion to the user, based on the evaluation of the user's risk type. More specifically, a user yielding a high value for a high-risk propensity would receive a suggestion of a high-risk choice and vice versa, also on a scale from 0 to 9.
This suggestion was followed by an explanation of all ten variables and their relevance to the estimated risk type of the user. For example, being female was an indicator towards less risk propensity, whereas being male was an indicator for a high-risk propensity. We analyzed the HRV and Facial Expressions during theses episodes to detect arousal.
After this explanation, the user could revise his/her decision. In the last phase, the users were asked to (verbally) name the features that they remembered from the explanation. After this, they were presented the features one after another and were asked to determine whether a certain value of this feature was an indicator for higher or lower risk propensity. We used this information as a proxy for understanding.

\subsubsection{Questionnaire}

During the first phase of the interaction, the participants were asked to fill out an online questionnaire, with the virtual robot Flobi being present and asking the questions.
A total of ten questions were asked. We refer to these questions as the ``variables'' or ``features'' that the system uses to compute and explain the risk type of the participant.

Based on results from empirical studies reported in the literature for each variable, a scoring scheme was developed that assigned a score for each answer indicating higher or lower risk propensity. Overall, this resulted in a linear scoring scheme -- the more scores, the higher the risk propensity. Scores were assigned  for each feature based on a median value reported in the literature above or below which a higher (+1) or lower (0) risk score was given.
This resulted in a risk type value between 0 and 9 for each participant.

\subsubsection{Emotion Induction}\label{sec:ei}

To investigate the effect of prior non-task related emotions on retention and understanding, we induced emotions via a biographic event recall similar to the one used in \cite{thommes2024xai}. The participants were randomly assigned to one of the two emotions: Fear and Happiness. For the induction, they were asked to remember an actual event where they experienced fear (or happiness). They were given 5 minutes time to mentally replay the situation to induce the emotion.

\subsubsection{Risk decision: First choice}

Directly after the emotion induction, the participants were explained the risk task and asked to provide their first decision. This first decision is important information to determine if the DSS' suggestion and explanation has an effect on the participant's (second) decision. That is, if the participant changes her selection in the direction of the system's suggestion, this is an indicator of advice taking.

\subsubsection{XAI decision and explanation}

\begin{figure}[H]
    \centering
    \includegraphics[width=1\linewidth]{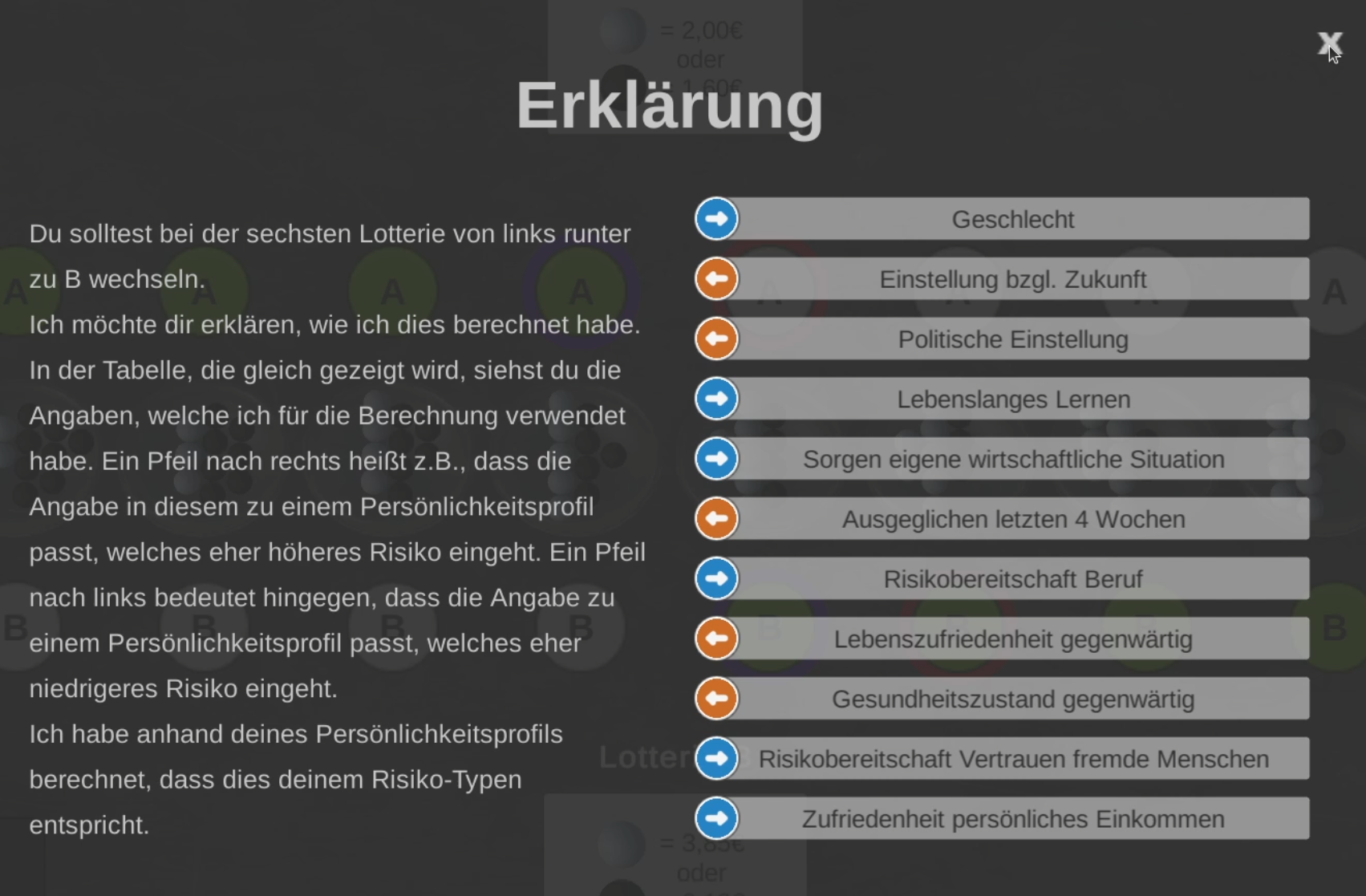}
    \label{fig:visualExplanation}
    \caption{Visual representation while Flobi was explaining which variables contributed to which risk type classification of the explainee. The red arrows indicated that the explainee's value of this variable contributed to an estimation of a lower risk type whereas a blue arrow indicated evidence for a higher risk type.}
\end{figure}

In the explanation phase (phase 4, see above), Flobi would explain for each variable, whether the explainee's (self-rated) value (e.g., of her political orientation) was an indicator adding to a higher or lower probability of the explainee as being more risk friendly. In this case, Flobi would say: ``Because your political orientation is rather left, you are presumably less risk friendly''. Fig. \ref{fig:visualExplanation} shows the GUI, where all variables are depicted with their respective contributions to the AI system's decision. This would be repeated for all 11 variables.

\subsubsection{Assessment of user's retention and understanding of the explanations and the task}

\begin{figure}[H]
    \centering
    \includegraphics[width=1\linewidth]{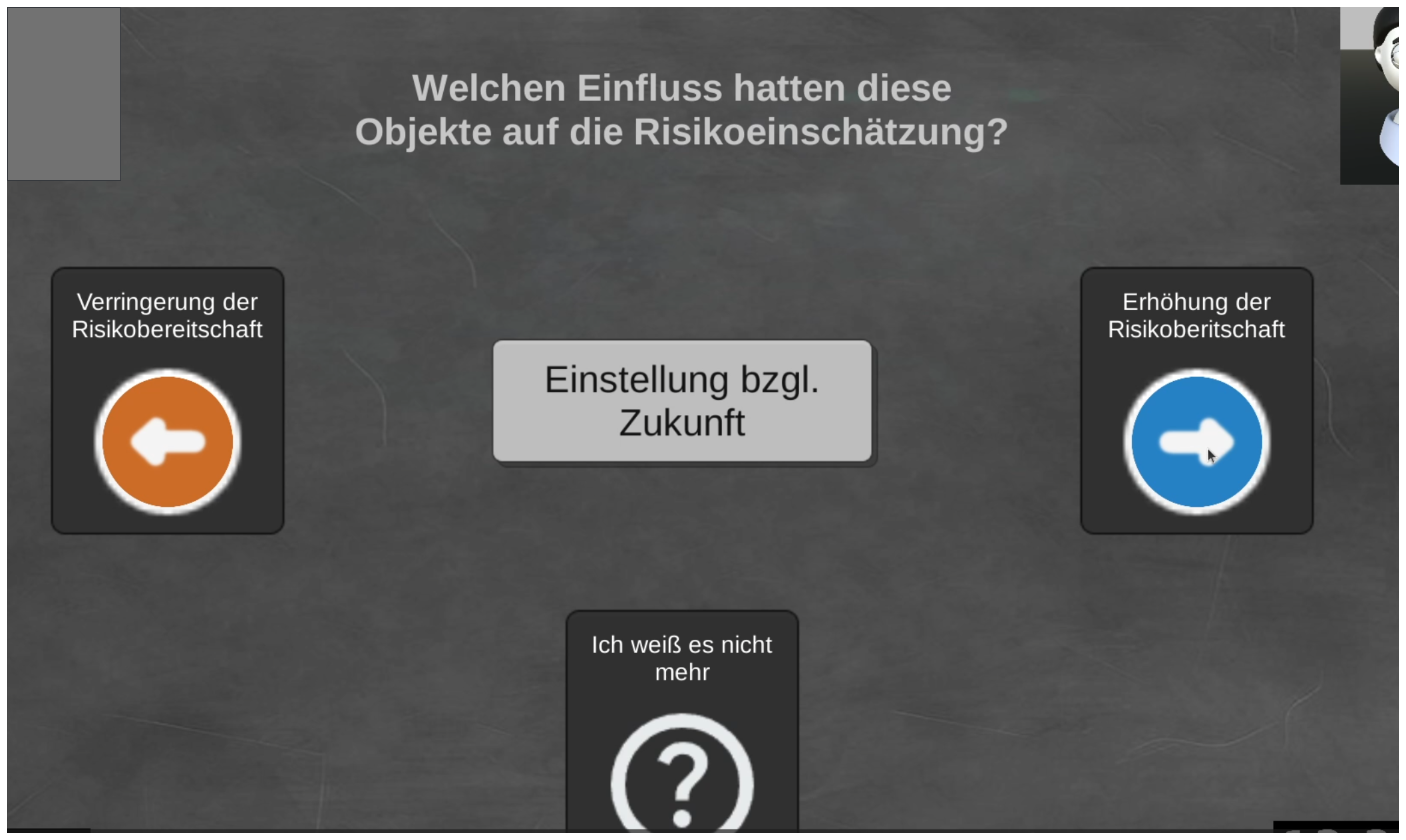}
    \caption{Visualization of the user interface to answer the question what effect the user's value of the presented variable had on the system's estimation of the user's risk type. This task was used to measure the explainee's retention of each variable.}
    \label{fig:GUIRecallQuestion}
\end{figure}

In the last phase, the uers' retention and understanding were assessed using the procedures described in subsection \ref{sec:DepVar}. See Fig. \ref{fig:GUIRecallQuestion} for the GUI to sort the features according to the remembered influence it had on the risk type classifcation. 


\section{Results}

\subsection{Emotion Manipulation Check}

\begin{figure}
    \centering
    \includegraphics[width=1\linewidth]{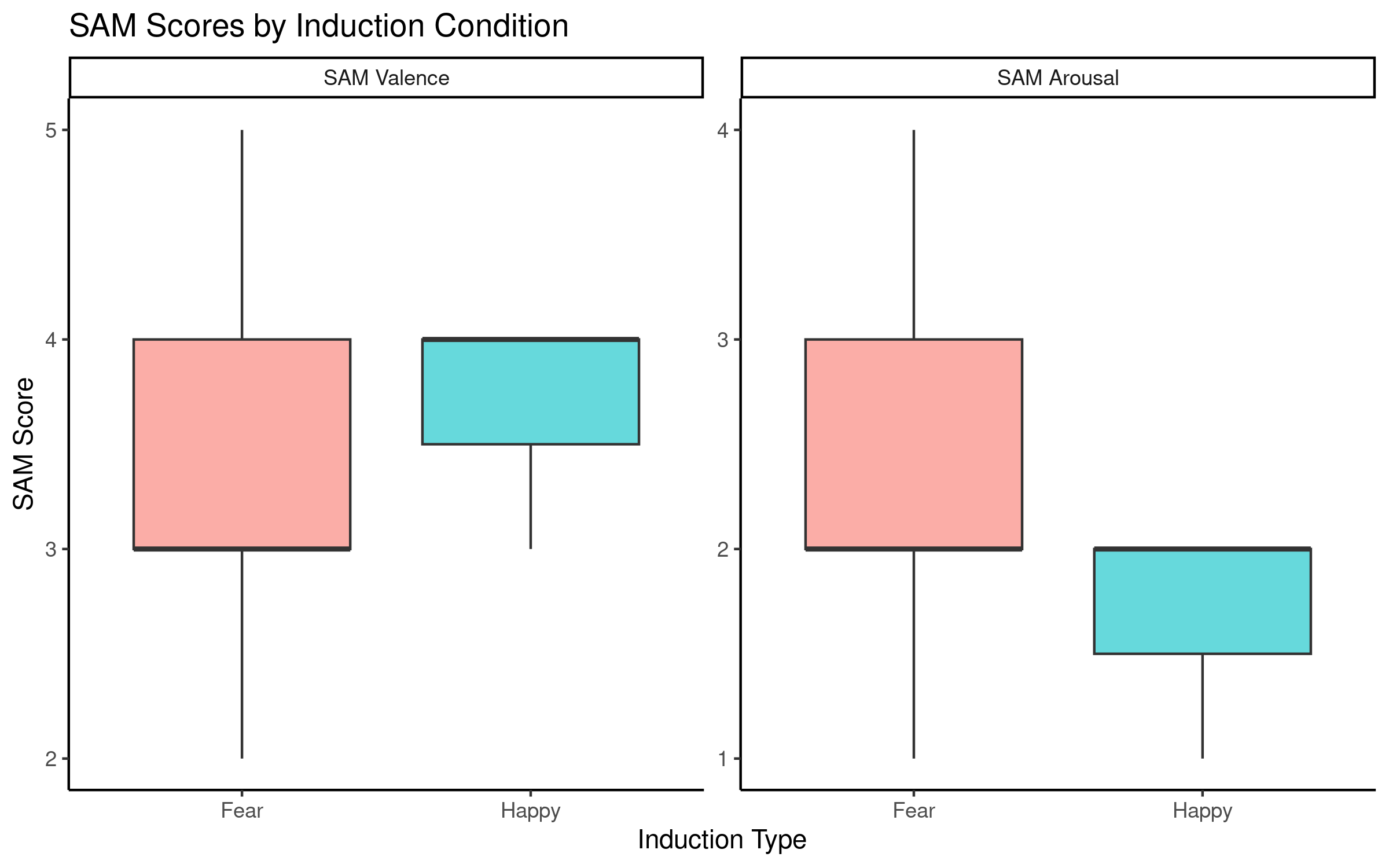}
    \caption{SAM scores for Valence (left) and Arousal (right), indicating higher (= more positive) valence for participants in the Happy condition, and a trend to higher self-reported arousal for participants in the Fear condition.}
    \label{fig:emotion-check}
\end{figure}

To examine differences in affective responses between induction conditions, we compared SAM Valence and Arousal scores across groups (Fear vs. Happy - cf. Fig. \ref{fig:emotion-check}).
On the Valence scale, which ranges from 1 (“unpleasant feeling”) to 5 (“pleasant feeling”), participants in the Happy condition reported higher valence ratings (median $\approx$ 4), indicating more pleasant emotional states compared to the Fear group (median $\approx$m 3).
Similarly, on the Arousal scale (1 = “calm”, 5 = “aroused”), the Fear group tended to report higher arousal levels than the Happy group, suggesting that the fear induction elicited more physiologically activating emotional responses.
These trends align with theoretical expectations -- fear typically evokes high arousal and negative valence, while happiness is associated with positive valence and lower arousal.

\subsection{Influence of task-unrelated prior emotion on retention}
To answer the \textbf{RQ1a} (Do task-unrelated emotions influence the retention of explained feature relevance?) the mean recall of the verbally and visually explained features was measured for each condition by analyzing the verbal answer to the question which features the user remembered. Note that this was an open question requiring the users to actively retrieve and formulate the name of the features while ignoring the meaning it had on the outcome. Since we recorded the participant's voice during the whole experiment, we were able to capture their comments also during the explanation of the features. As noted above, due to a programming error, one feature (``Einstellung bzgl. Zukunft'' - attitude towards future) was reported wrongly for the majority of participants. In these cases, some participants would comment the mistake spontaneously through a verbal utterance. However, some participants also commented features that were communicated correctly. For the participants, there was no difference between these cases -- they experienced both cases as a mistake from the system. As these comments indicated an epistemic reaction (surprise) we also investigated their effect on retention.

\begin{figure}
    \centering
    \includegraphics[width=0.45\linewidth]{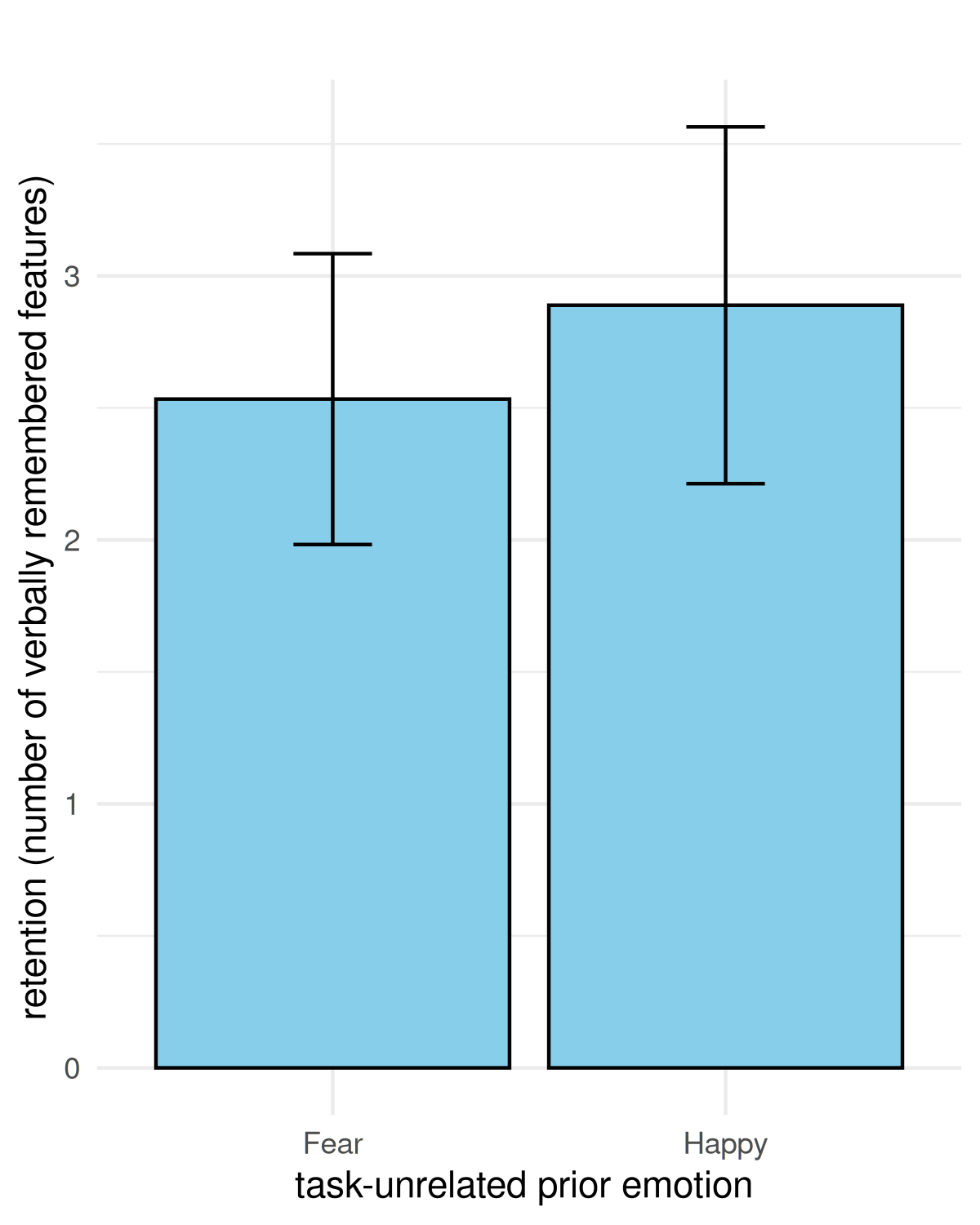}
    \includegraphics[width=0.45\linewidth]{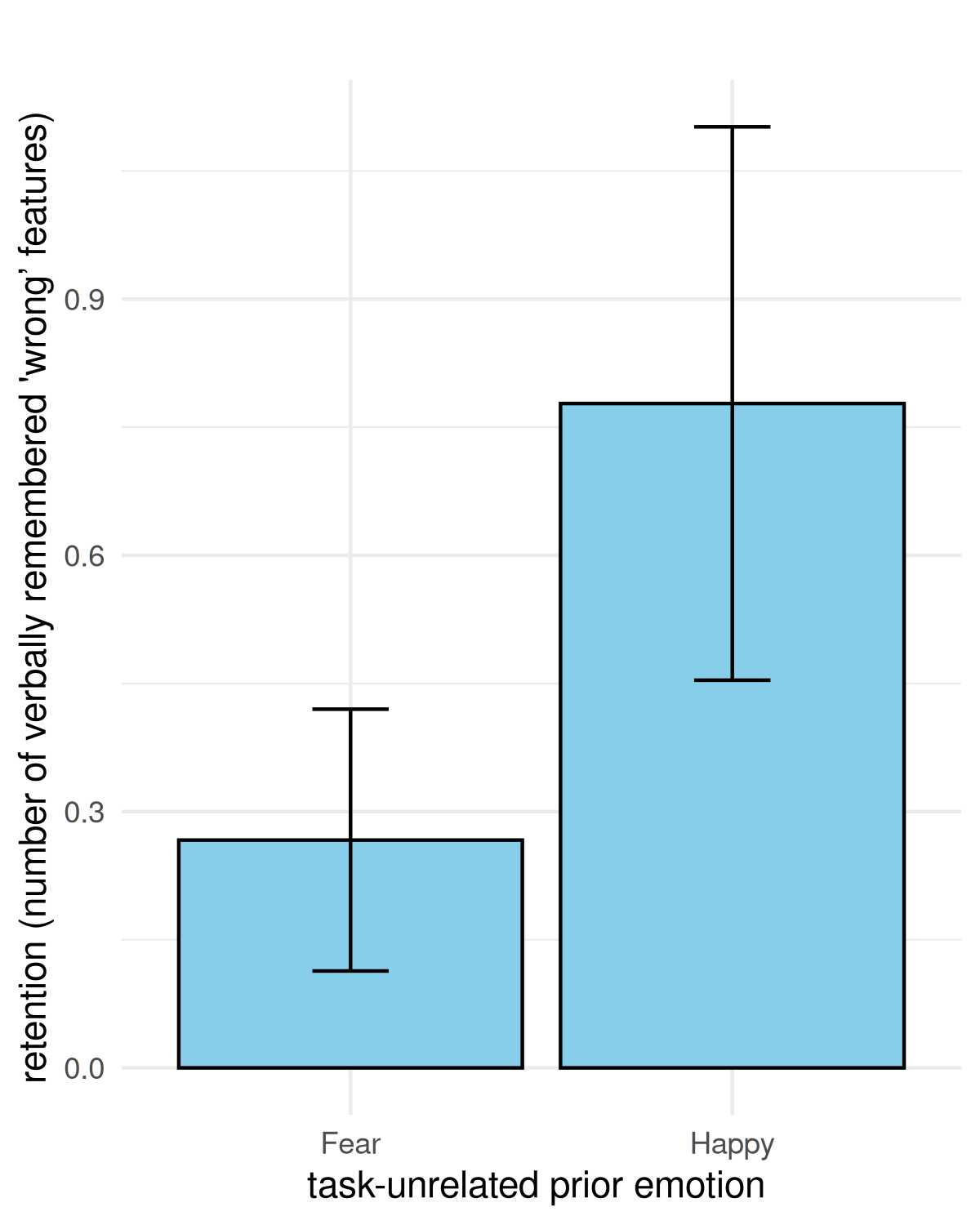}
    \caption{(left) Mean feature retention divided for task-unrelated prior emotion. (right) Mean retention of 'wrong' feature divided for task-unrelated prior emotion.}
    \label{fig:reattention_induction}
\end{figure}
Figure \ref{fig:reattention_induction} shows the \textbf{retention} for each task-unrelated induced emotion (left). The right side shows the average verbal labeling of features that participants explicitly identified as incorrect. 
A one-way ANOVA was conducted to compare the mean retention between the two induced prior emotions (Fear, Happiness). The analysis revealed no significant difference $(F(1, 22) = 0.1619, p = .6913)$. Thus, the retention of features is not influenced by a prior emotion.

However, those feature explanations that were perceived as erroneous by some participants may have yielded a better retention as they attracted attention.
To investigate if the induced emotion had an effect on the perceived incorrect features,
a one-way ANOVA was conducted to examine the effect of the induction condition on the number of recall of the perceived incorrect features. The analysis revealed no significant main effect of induction, $F(1, 22) = 2.59, p = .122$. However, as can be seen in Fig. \ref{fig:reattention_induction} participants in the condition ``Happy'' remembered almost twice as many features they had commented as wrong as those in the ``Fear'' condition.


\subsection{Influence of task-unrelated prior emotion on understanding}\label{sec:ei_understanding}


Addressing \textbf{RQ1b} (Do task-unrelated emotions influence the understanding of explained feature relevance?), 
the explainees were asked to indicate for each variable which influence it had \textbf{in their own case} on their risk propensity as estimated by the AI system (see Fig. \ref{fig:GUIRecallQuestion}). 

\begin{figure}
    \centering
    \includegraphics[width=1\linewidth]{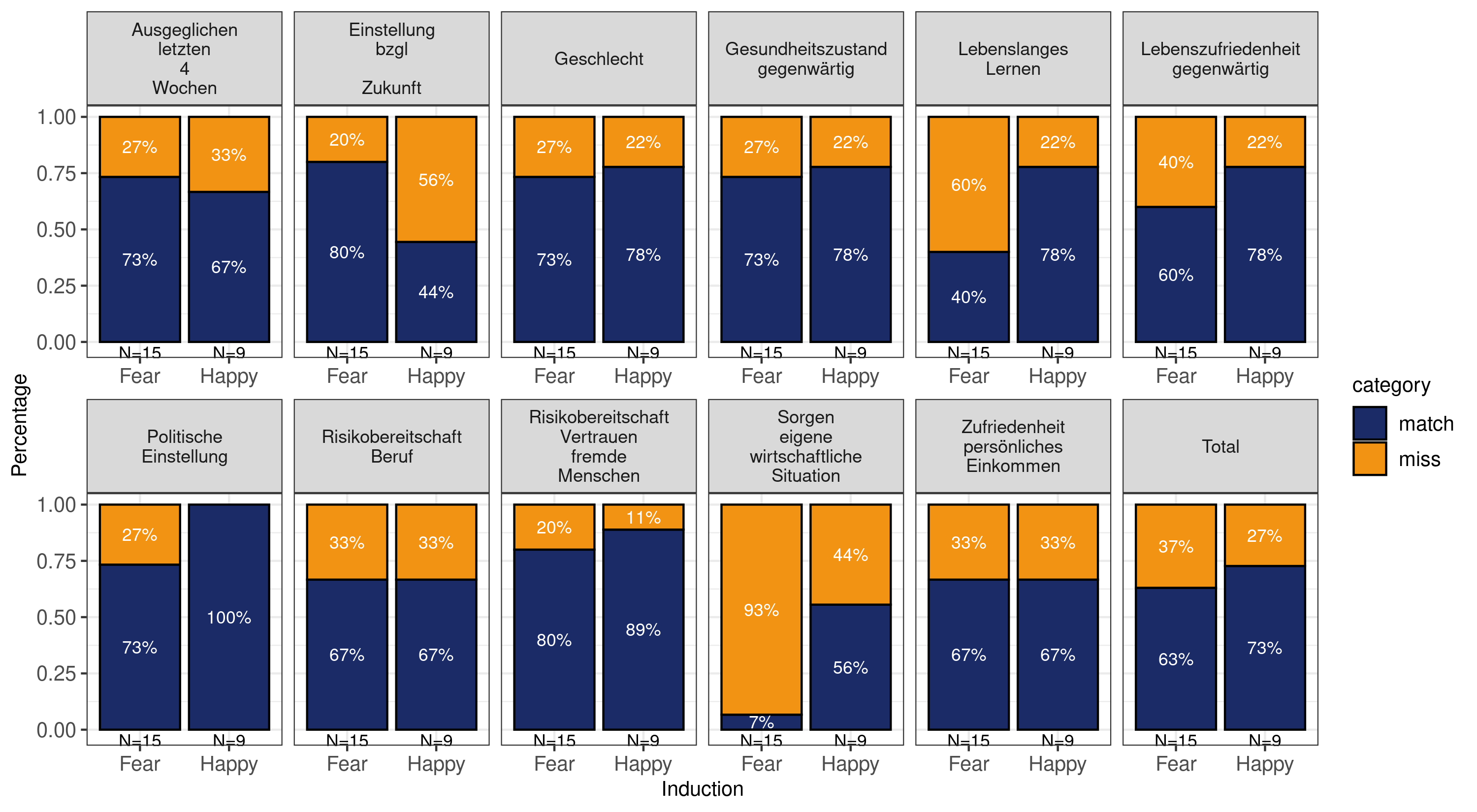}
    \caption{A comparison of the understanding of each presented feature, as tested by the recall task shown in Fig. \ref{fig:GUIRecallQuestion},differentiated by task-unrelated induced emotion.}
    \label{fig:UnderstandingFeatureContribution}
\end{figure}

An ANOVA was conducted to examine the effects of task-unrelated emotions (\textit{emotion induction}) and explained \textit{feature} and their interactions with the outcome variable \textit{understanding}.

There was a significant main effect of the explained feature, $F(9, 220) = 3.51, p < .001$, suggesting that the feature categories differed significantly in their association with the outcome \textit{understanding}.
The main effect of \textit{emotion induction} was marginally significant, $F(1, 220) = 3.23, p = .074$, indicating a potential trend for the induction condition, i.e. the task-unrelated emotion, to influence responses.
The interaction effect \textit{induction emotion } × \textit{explained feature} interaction was not statistically significant, $F(9, 220) = 1.62, p = .111$.

For a more qualitative analysis, we visualized the matching (``match'') vs. non-matchin (``miss'') answers of the participants with the explantation they had received in the previous phase from Flobi (see Fig. \ref{fig:UnderstandingFeatureContribution}). 
Interestingly, we see that the feature that was explained wrongly to most of the participants (``Einstellung bzgl. Zukunft'') was remembered differently by the participants with different induced emotions.
While 80\% of the participants of the Fear condition agreed with Flobi's (wrong!) explanation, only 44\% of the participants of the Happy condition did so. This is somewhat surprising, as one would expect that fear is generally associated with suspicion, leading to scrutinizing offered information and suggestions. Yet, in this case, it seems that participants in the Happy condition remembered their own decision better. 


A different interpretation might be, that there are two effects of emotion on retention: (1) Users do not remember their answer to these specific questions about an uncertain future correctly, as they were given in a neutral state. Rather, they ``remember'' their current emotionally tainted attitude towards the future: in the condition Fear this would be negative, in the condition Happy this would be positive. Indeed, in the ATF it is being argued that happiness gives  high attribution to certainty, fear towards low certainty.
(2) Users judge the impact of these variables according to their current emotion: users in the Fear condition believe that being uncertain about the future reduces the risk  propensity (although research shows a different relationship: high uncertainty about the future increases risk propensity, low uncertainty decreases it), whereas users in the Happy condition believe that being uncertain about the future increases risk propensity. This might be seen as some kind of confirmation bias or transfer, as this estimated influence on risk propensity corresponds to their own risk propensity at that time: being happy (rather than being certain about the future) increases risk propensity, whereas being fearful (rather than being uncertain about the future) decreases risk propensity.

Although it is not certain, what exactly causes these different judgment results, it is clear that the emotional state can affect how the influence of certain variables on the outcome is remembered or judged. 
Yet, it remains unclear how such an effect could be detected in interaction with an intelligent, explainable AI system.


\subsection{Effect of explanations on arousal}

To determine the effect of explanations on arousal,
we defined the emotional reactions -- or arousal -- by using the Emonet arousal data during the feature presentation in combination with the rolling z‑score with k = 2.5 and a window of 500ms (cf. equations (1) and (2)). In this way, we determined the peaks of arousal that stood out from the preceding arousal values, indicating emotional reactions.
Fig. \ref{fig:emotionalReaction-label} shows the arousal values as computed by EmoNet as a red line, plotted over time, for participant 25.
The feature names (rotated vertically) denote the beginning of the verbal (and visual) explanation of the according feature. For example, the explanation of the feature ``Geschlecht'' (``gender'') starts at the second 0.

Vertical black lines indicate a positive z-score and therefore an emotional reaction. For example, the black line at second 4 indicates an arousal bout right at the beginning of the feature explanation for ``Politische Einstellug'' (``political attitude'').
The yellow lines indicate a rapidly downfall of the measured arousal.

\begin{figure}
    \centering
    \includegraphics[width=1\linewidth]{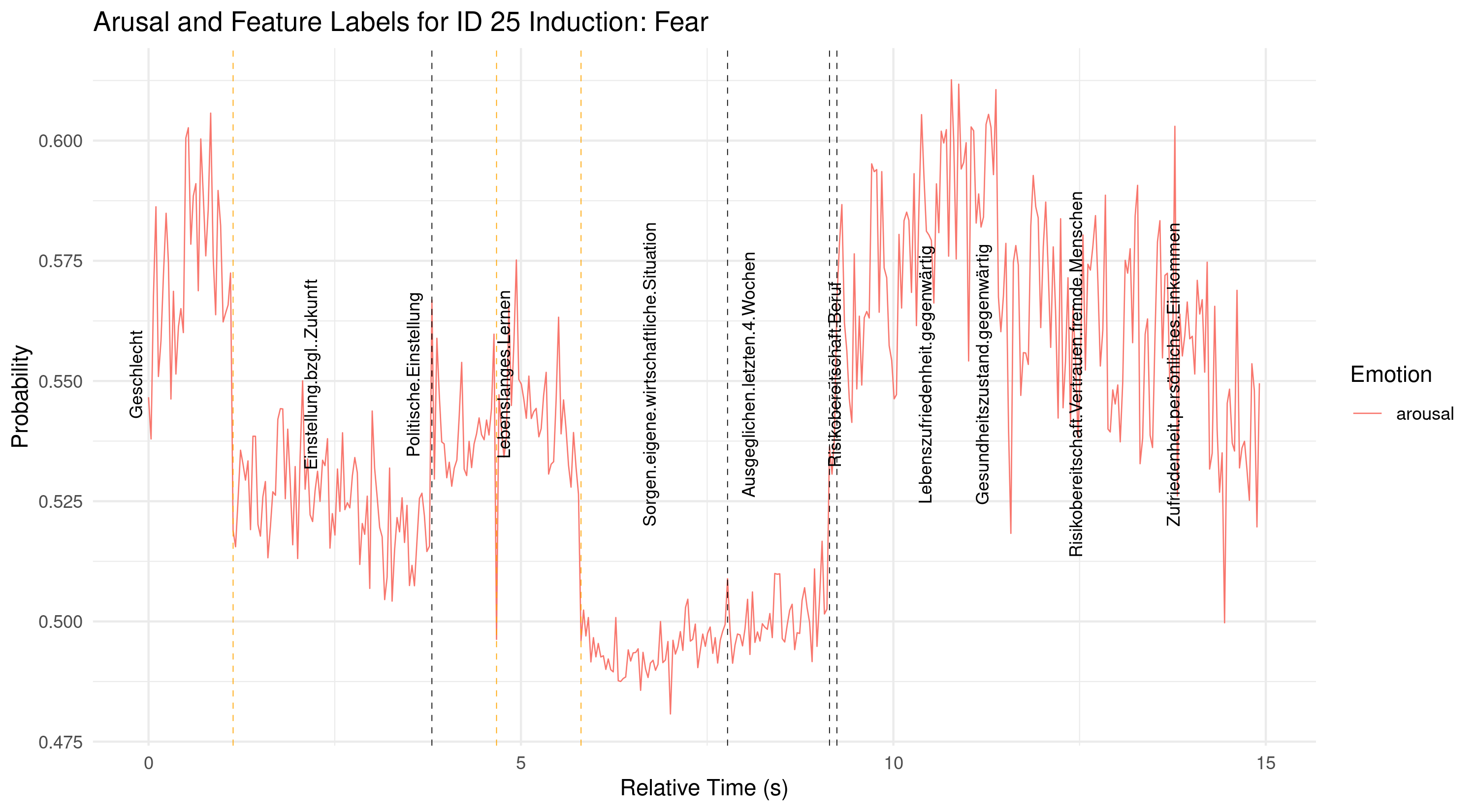}
    \caption{Arousal over the Feature Presentaion Time with emotional reaction detection}
    \label{fig:emotionalReaction-label}
\end{figure}

If a feature explanation segment contained one (or more) bouts of arousal -- as indicated by the black lines -- it was counted as a feature explanation causing an emotional reaction (or arousal).

\begin{figure}
    \centering
    \includegraphics[width=1\linewidth]{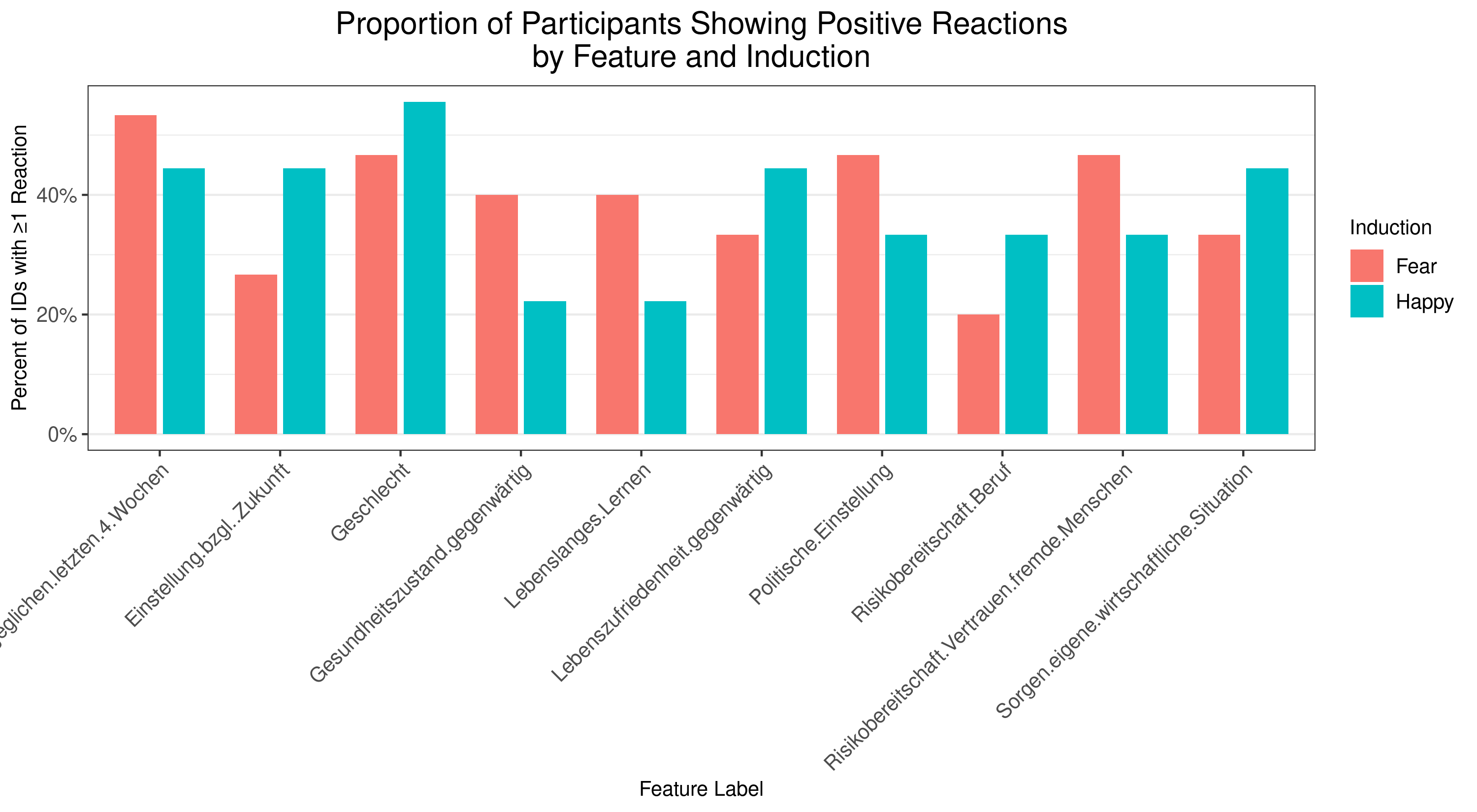}
    \caption{Enter Caption}
    \label{fig:EmotionalReactionByFeature}
\end{figure}

Fig. \ref{fig:EmotionalReactionByFeature} shows the proportion of participants per feature for whom arousal was measured. 
As can be seen, there are differences in frequency of arousal for the different features. The features ``gender'' (``Geschlecht'') and ``even temper of the last 4 weeks'' (``Ausgeglichenheit...'') yielded most frequent bouts of arousal with almost 60\% of the participants whereas ``risk propensity job'' (``Risikobereitschaft Beruf''), ``livelong learning'' (``lebenslanges Lernen''), and ``current health'' (``Gesundheitszustand gegewaertig'') yielded least frequent arousal with about 20\%.
This indicates that features differ in their potential to evoke arousal. What the underlying reasons for the arousal are, remains unclear, so far. But there may be intrinsic (e.g., the personal relevance of this feature for each participant) as well as extrinsic (e.g. recency effect, length of word / ease of word) reasons. 

In addition to these differences we also see different frequencies of arousal between participants in the Fear and the Happy group. 
Most striking is the difference in the feature  ``attitude towards future'' (``Einstellung bzgl. Zukunft'') which raised arousal in about 25\% of the participants in the Fear group compared to about 45\% in the Happy group. Most strikingly, it was this last feature where an error occurred. Thus, arousal would be expected in both groups. However, this
cannot be observed in the Fear group, from where arousal would be expected most frequently. 

Thus, overall we see that explanations can induce arousal. However, so far no clear pattern as to what factors actually cause the arousal is recognizable.

\subsection{Effect of explanation-induced arousal on retention} 

\begin{figure}
    \centering
    \includegraphics[width=0.45\linewidth]{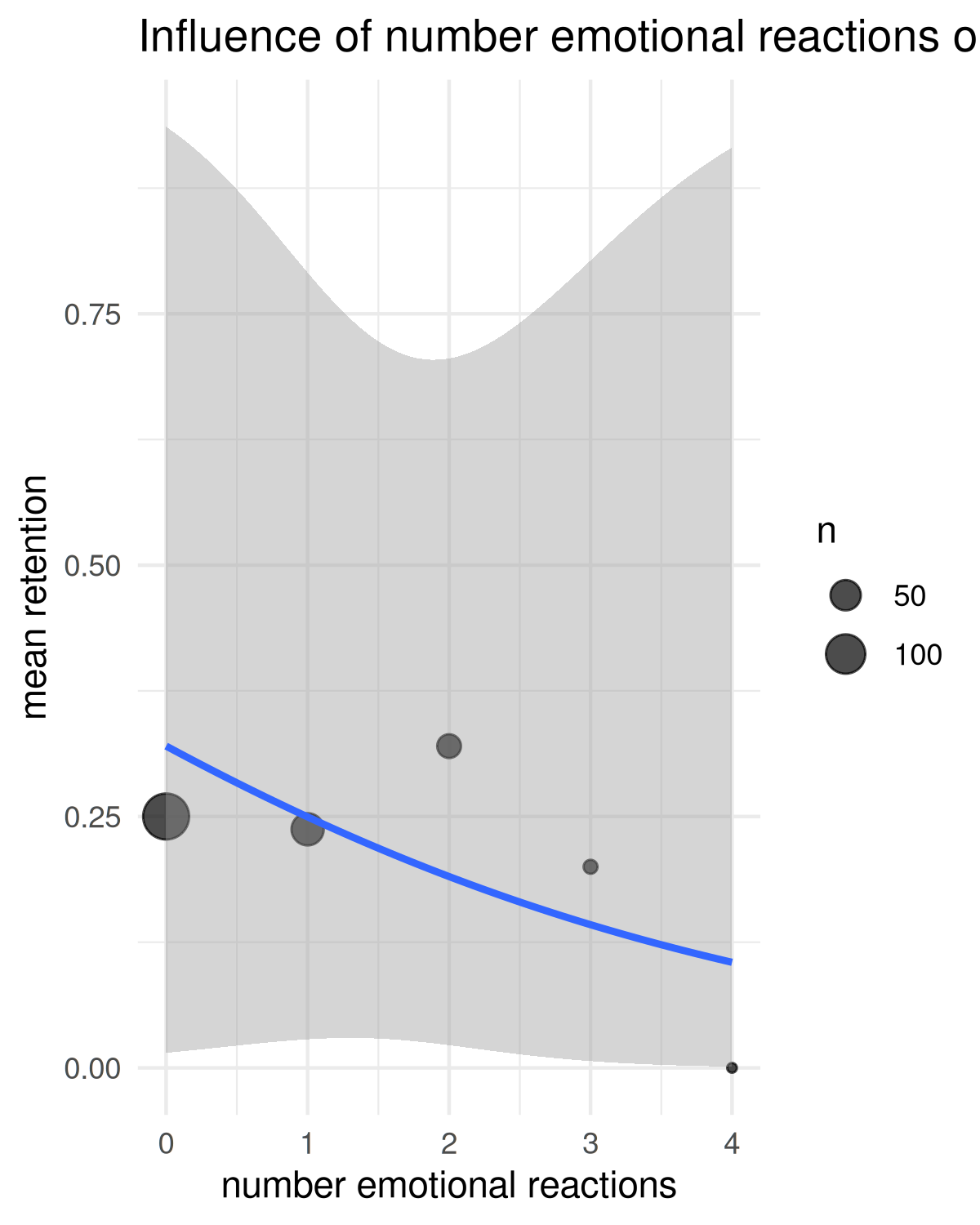}
      \includegraphics[width=0.45\linewidth]{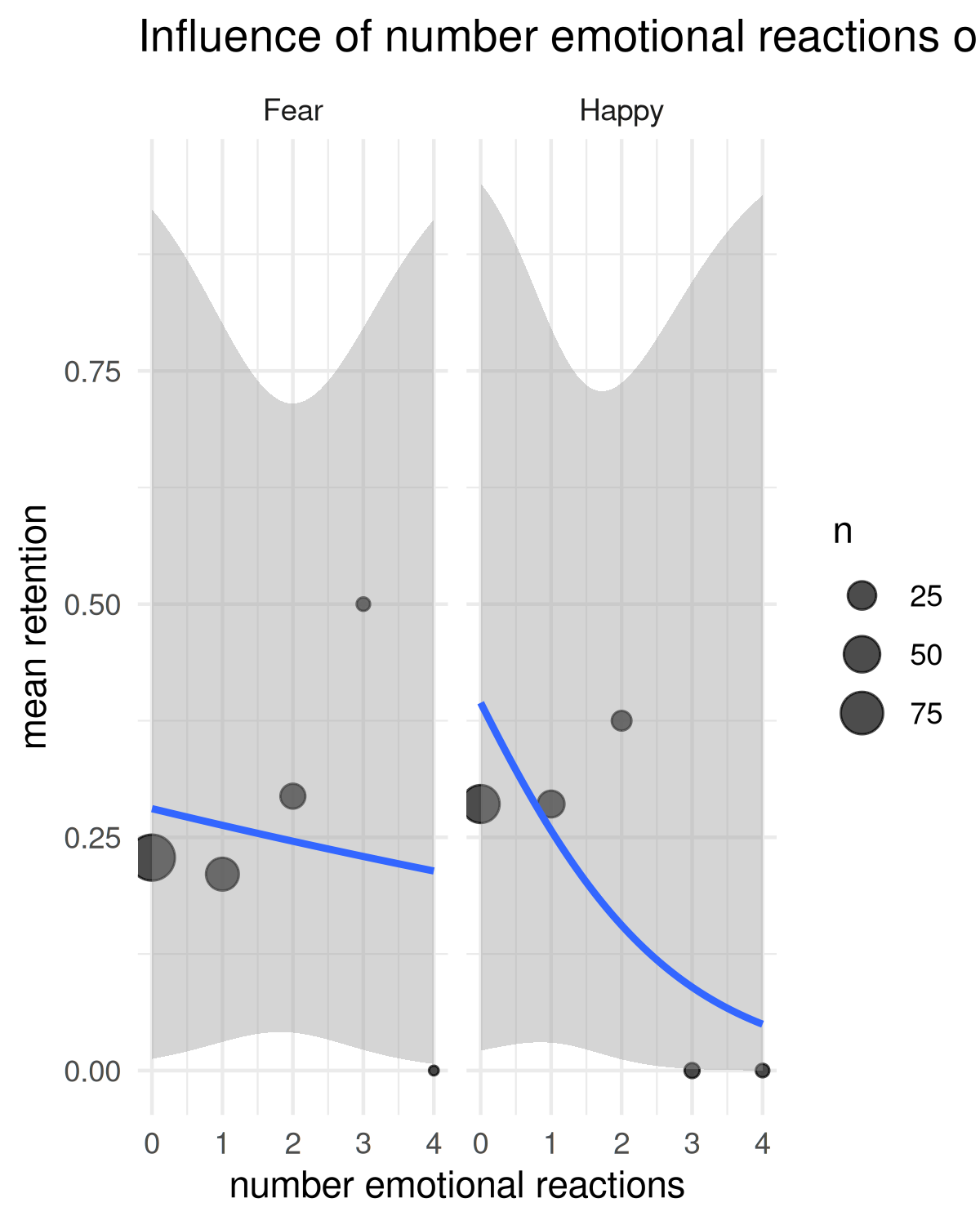}
    \caption{The mean retention of the feature divided on number of emotional reactions during the feature presentation for all (left) and divided by condition (right).}
    \label{fig:er_retantion}
\end{figure}

Figure \ref{fig:er_retantion} visualizes the mean retention as a function of the number of emotional reactions that a feature explanation evoked. The size of the bullet visualizes the number of occurrences. Here, the explanations of all 10 features for all 24 participants (10 x 24 = 240) have been taken into account. For example, in the left figure the largest dot a $number emotional reactions = 0$ indicates that the mean retention for those 100 (given by the size of the dot) feature explanations that yielded 0 emotional reactions (i.e. bouts of arousal) was 25
On the right side the graph is split into the reactions of the participants from the Fear and the Happy condition.

To examine whether emotional reactions were associated with participants’ retention of individual features, we fitted a generalized linear mixed model (GLMM) with a binomial distribution and logit link to predict the binary outcome \textit{retention}. 
The fixed effects included \textit{emotional reactions (i.e. arousal)}, \textit{emotion induction (i.e. Happy vs Fear)}, and the \textit{explained feature}, while a random intercept was included for participant ID (N = 240 observations, 24 participants).
The binary dependent variable indicated whether a feature was verbally recalled (retention = 1) or not (retention = 0). 
Model estimation was performed using the glmer() function from the lme4 package in R.

The predictors \textit{emotional reactions}, \textit{emotion induction}, did not reach significance (p > .10).
This means that neither arousal nor the induced emotion had an influence on the recall of a feature.

The model showed a significant negative intercept ($\beta = -2.95, SE = 0.89, z = -3.33, p < .001$), suggesting a low baseline probability of the retention of all features.
Among the fixed effects, the following explained feature variables were significant predictors and more likely to recall verbally:
\begin{itemize}
    \item Gender: $\beta = 3.11, SE = 0.93, z = 3.34, p < .001$
    \item Current health status: $\beta= 2.87, SE = 0.92, z = 3.11, p = .002$
    \item Political orientation: $\beta = 2.67, SE = 0.92, z = 2.90, p = .004$
\end{itemize}

This means that the features Gender, Current health status, and Political orientation are predictors for retention. This is an interesting result as Gender was the feature with the most frequent bouts of arousal, whereas Current health status yielded the least frequent bouts of arousal which indicates that arousal alone may not be a good predictor of retention.



\subsection{Effect of explanation-induced arousal on understanding} 


While we were not able to show a significant effect of arousal on retention it is possible that arousal affects understanding. We applied the same approach as for retention and fitted a generalized linear mixed model (GLMM) with a binomial distribution and logit link to predict the binary outcome \textit{understanding}. The fixed effects included \textit{emotional reactions}, \textit{emotion induction}, and the \textit{explained feature}, while ID was modeled as a random intercept to account for individual variability.
Model estimation was performed using the glmer() function from the lme4 package in R.

The model showed a significant positive intercept ($\beta = 1.07, SE = 0.52, p = .040$), indicating a relatively high baseline probability of the outcome \textit{understanding}. That is, the participants had a good understanding of what effect each feature had on the outcome of the system.

The \textit{emotional reaction} is significantly negatively associated with the outcome ($\beta = -0.45, SE = 0.19, p = .015$), suggesting that higher levels of positive\textit{ emotional reactions} were associated with lower \textit{understanding}. Thus, too much arousal or too many bouts of arousal hinder understanding.

The effect of the \textit{emotion induction}  was not significant ($\beta = 0.64, p = .108$), indicating no clear effect of the induced emotions Fear and Happiness on the outcome in this model.

Among the \textit{explained features}, only  "Concerns about one's own economic situation" had a significant negative effect ($\beta = –2.31, p < .001$) on the \textit{understanding}.

The random intercept for ID had a variance of 0.28 (SD = 0.53), indicating some variability in baseline response tendencies across participants.
Thus, we see individual effects on the capability to understand the meaning of the effect of a feature on the outcome of the DSS.

\begin{figure}
    \centering
    \includegraphics[width=0.45\linewidth]{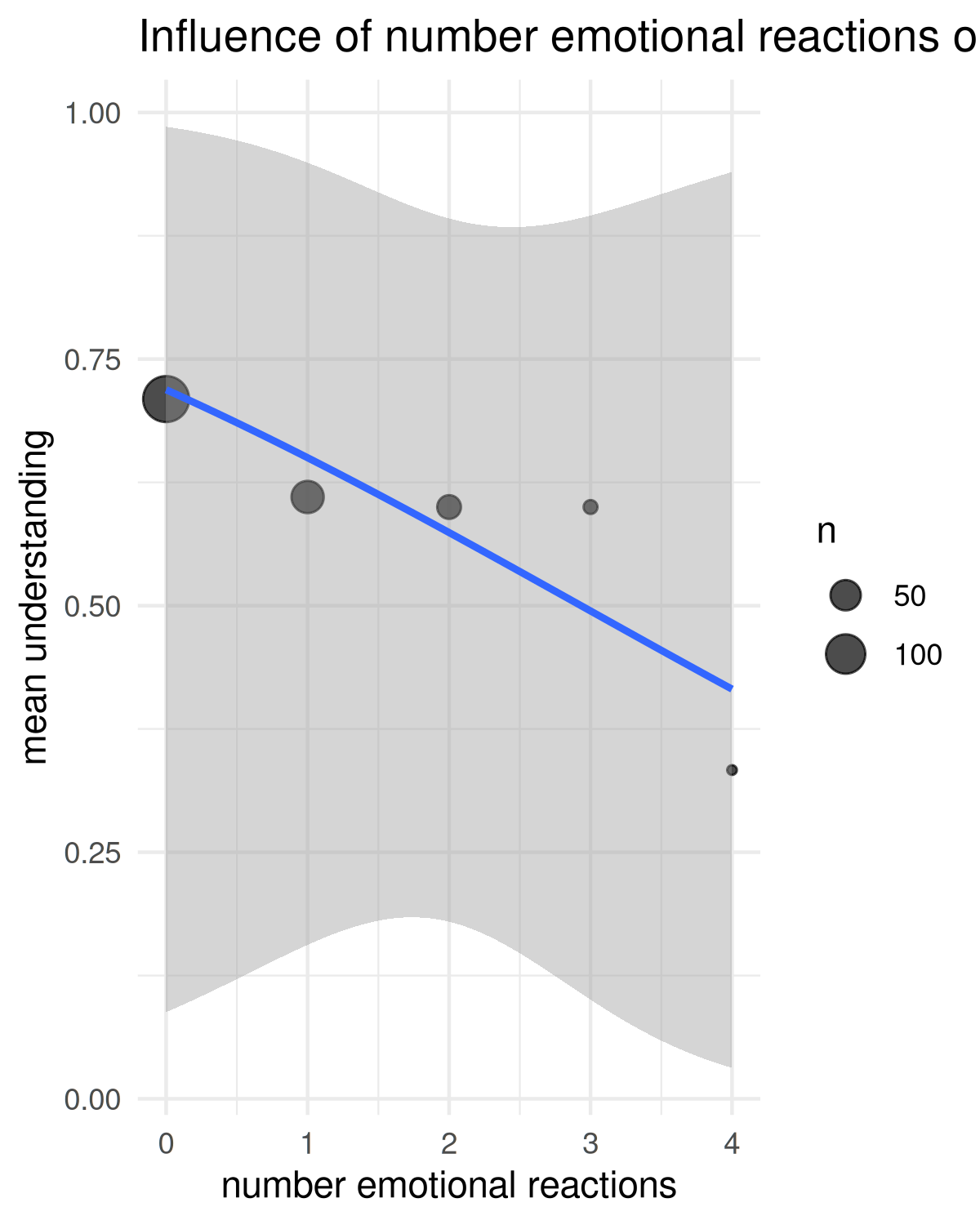}
    \includegraphics[width=0.45\linewidth]{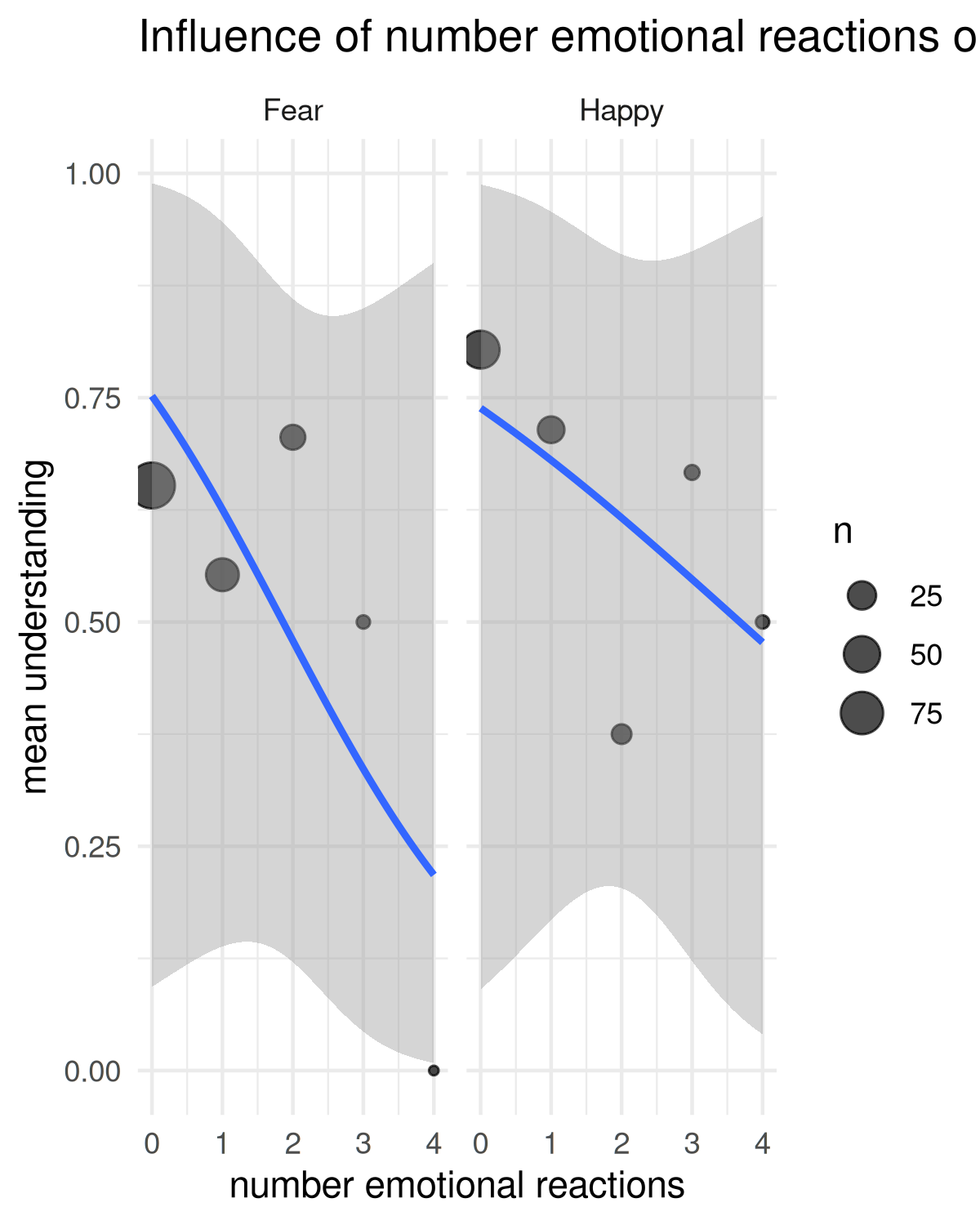}
    \caption{The mean understanding of the feature divided for number of emotional reactions during the feature presentation for all (left) and divided by condition (right).}
    \label{fig:emotionalReaction-label}
\end{figure}

\section{Discussion}

In the following, we will discuss our results regarding our initial research questions.

\textbf{RQ1a:} Do task-unrelated emotions influence the retention of explained feature relevance?
\textbf{RQ3a:} Do task-related emotions influence retention of the explanation features?

Our results indicate that neither task-unrelated prior emotions nor task-generated emotional reactions (or arousal) were significant predictors of retention. That is, although there is evidence that a certain amount of arousal in general improves retention of information, this was not the case in the current study. There are many possible explanations for this. One explanation might be, that the complex explanation situation was novel and required a high cognitive load due to the new way of explaining the relevance of features. 
Another explanation might be that the feature variables themselves may not have been understood by the participants. Some variable names are very long and might be difficult to remember, so that participants were not able to map certain variable or feature names to the questions they had been asked in the initial questionnaire. This would require an addition explanation layer that allows the participant to ask for an explanation concerning the different (or globally most relevant) features.

\textbf{RQ2:} Which features trigger emotional reactions during explanation? (task-related emotions)

Our results indicate that certain individual characteristics -- such as gender, current health status, and political orientation -- are significant predictors for the recall of explained features.
Further research needs to investigate what characteristics render features so salient that they are remembered better than others.


\textbf{RQ1b:} Do task-unrelated emotions influence the understanding of explained feature relevance?

We found a main effect of the explained feature on understanding but only a marginal effect that task-unrelation emotions influence understanding.
This indicates that the features differed significantly in how well they were understood by the participants. However, the influence of task-unrelated emotions on understanding was only marginal.


\textbf{RQ3b:} Do task-related emotions influence understanding of the explanation features?

Most interestingly, we found that the emotional reaction, i.e., arousal, is significantly negatively associated with understanding, suggesting that higher levels of positive emotional reactions were associated with lower understanding. This is in accordance with other findings that indicate that too much arousal (or too many bouts of arousal, in our case) hinders understanding. Thus, it is an important goal to finde the right amount of arousal in order to foster understanding in a DSS scenario.




\begin{ack}
This research was funded by the Deutsche Forschungsgemeinschaft
(DFG, German Research Foundation): TRR 318/1 2021-438445824 “Constructing Explainability”.
\end{ack}



\bibliography{References}

\end{document}